\documentclass[twocolumn]{aastex}
\usepackage{graphicx,floatrow,amsmath,gensymb,rotating}
\usepackage{subfigure}
\usepackage{natbib}
\usepackage{xcolor}
\usepackage{threeparttable,booktabs}
\usepackage{etoolbox}

\newcommand{\kms}{\>{\rm km}\,{\rm s}^{-1}}

\begin{document}

\title{The Active Galactic Nuclei in the Hobby-Eberly Telescope Dark Energy Experiment Survey (HETDEX) III. A red quasar with extremely high equivalent widths showing powerful outflows}

\author[0000-0001-5561-2010]{Chenxu Liu}
\affiliation{Department of Astronomy, The University of Texas at Austin, 2515 Speedway Boulevard, Austin, TX 78712, USA}
\correspondingauthor{Chenxu Liu}
\email{lorenaustc@gmail.com}

\author[0000-0002-8433-8185]{Karl Gebhardt}
\affiliation{Department of Astronomy, The University of Texas at Austin, 2515 Speedway Boulevard, Austin, TX 78712, USA}

\author[0000-0002-0417-1494]{Wolfram Kollatschny}
\affiliation{Institut f\"ur Astrophysik, Universit\"at G\"ottingen, Friedrich-Hund Platz 1, 37077 G\"ottingen, Germany}

\author[0000-0002-1328-0211]{Robin Ciardullo}
\affil{Department of Astronomy \& Astrophysics, The Pennsylvania State University, University Park, PA 16802, USA}
\affil{Institute for Gravitation and the Cosmos, The Pennsylvania State University, University Park, PA 16802, USA}

\author[0000-0002-2307-0146]{Erin Mentuch Cooper}
\affiliation{Department of Astronomy, The University of Texas at Austin, 2515 Speedway Boulevard, Austin, TX 78712, USA}
\affiliation{McDonald Observatory, The University of Texas at Austin, 2515 Speedway Boulevard, Austin, TX 78712, USA}

\author[0000-0002-8925-9769]{Dustin Davis}
\affiliation{Department of Astronomy, The University of Texas at Austin, 2515 Speedway Boulevard, Austin, TX 78712, USA}

\author[0000-0003-2575-0652]{Daniel J. Farrow}  
\affiliation{University Observatory, Fakult\"at f\"ur Physik, Ludwig-Maximilians University Munich, Scheiner Strasse 1, 81679 Munich, Germany}
\affiliation{Max-Planck Institut f\"ur extraterrestrische Physik, Giessenbachstrasse 1, 85748 Garching, Germany}

\author[0000-0001-8519-1130]{Steven L. Finkelstein}
\affiliation{Department of Astronomy, The University of Texas at Austin, 2515 Speedway Boulevard, Austin, TX 78712, USA}

\author[0000-0003-1530-8713]{Eric Gawiser}
\affiliation{Department of Physics and Astronomy, Rutgers, the State University of New Jersey, Piscataway, NJ 08854, USA}

\author[0000-0001-6842-2371]{Caryl Gronwall}
\affiliation{Department of Astronomy \& Astrophysics, The Pennsylvania State University, University Park, PA 16802, USA}
\affiliation{Institute for Gravitation and the Cosmos, The Pennsylvania State University, University Park, PA 16802, USA}

\author[0000-0001-6717-7685]{Gary J. Hill}
\affiliation{McDonald Observatory, The University of Texas at Austin, 2515 Speedway Boulevard, Austin, TX 78712, USA}
\affiliation{Department of Astronomy, The University of Texas at Austin, 2515 Speedway Boulevard, Austin, TX 78712, USA}

\author[0000-0002-1496-6514]{Lindsay House}
\affiliation{Department of Astronomy, The University of Texas at Austin, 2515 Speedway Boulevard, Austin, TX 78712, USA}

\author[0000-0001-7240-7449]{Donald P. Schneider}
\affiliation{Department of Astronomy \& Astrophysics, The Pennsylvania State University, University Park, PA 16802, USA}
\affiliation{Institute for Gravitation and the Cosmos, The Pennsylvania State University, University Park, PA 16802, USA}

\author[0000-0001-6746-9936]{Tanya Urrutia}
\affiliation{Leibniz Institut für Astrophysik, Potsdam, An der Sternwarte 16, 14482 Potsdam, Germany}

\author[0000-0003-2307-0629]{Gregory R. Zeimann}
\affiliation{Hobby-Eberly Telescope, University of Texas, Austin, Austin, TX, 78712, USA}

\begin{abstract}

We report an Active Galactic Nucleus (AGN) with extremely high equivalent width (EW), $\rm EW_{Ly\alpha+N\,V,rest}\gtrsim921\,\AA$ in the rest-frame, at $z\sim2.24$ in the Hobby-Eberly Telescope Dark Energy Experiment Survey (HETDEX\null) as a representative case of the high EW AGN population. The continuum level is a non-detection in the HETDEX spectrum, thus the measured EW is a lower limit. The source is detected with significant emission lines ($>7\sigma$) at Ly$\alpha+$\ion{N}{5} $\lambda1241$, \ion{C}{4} $\lambda1549$, and moderate emission line ($\sim4\sigma$) at \ion{He}{2} $\lambda1640$ within the wavelength coverage of HETDEX (3500\,\AA\ - 5500\,\AA\null). The $r$-band magnitude is 24.57 from the Hyper Suprime-Cam-HETDEX joint survey with a detection limit of $r=25.12$ at $5\sigma$. The $\rm {Ly\alpha}$ emission line spans a clearly resolved region of $\sim10\arcsec$ (85\,kpc) in diameter. The $\rm {Ly\alpha}$ line profile is strongly double peaked. The spectral decomposed blue gas and red gas Ly$\alpha$ emission are separated by $\sim1\farcs 2$ (10.1 kpc) with a line-of-sight velocity offset of $\sim1100\,\kms$. This source is probably an obscured AGN with powerful winds.

\end{abstract}


\section{Introduction}
\label{sec_intro}

Active Galactic Nuclei (AGN) are among the most energetic phenomena known in the universe. They can be identified with various observational techniques in different bands. Radio selection is sensitive to the powerful jet of luminous radio-loud AGN. X-ray has strong penetration and low dilution from the host galaxies, thus is the most efficient way in identifying low luminosity AGN \citep[e.g.,][]{Xue2016,Luo2017}. Mid-IR can identify both obscured and un-obscured AGN \citep[e.g.,][]{Stern2012}. However, different bands may represent different phases of AGN evolution and a single band identified AGN sample can be contaminated with strong starbursts. Deep UV/optical observations, especially the spectroscopic ones, are still a unique window to study the evolution of AGN.

Traditional optical spectroscopic surveys of AGN usually select targets based on photometric observations, i.e., detections in the broad band imaging with point-like morphologies to be distinguished from extended nearby galaxies, and blue colors to characterize the power-law continuum shape \citep[e.g.,][]{Richards2006}. The photometric selection can produce a selection effect on the AGN sample that fails to include the potential AGN population with optically faint host galaxies and high equivalent widths (EWs). These optically faint AGN could either be intrinsically less massive, which could make them outliers in the super-massive black hole (SMBH) - host relation \citep[see][for a review of the ``co-evolution'' between SMBH and their host galaxies]{Kormendy2013}, or 
red and obscured AGN.

Whether the co-evolution stands for all SMBHs and their host galaxies remains an open question. Early works in the 1990s reported the discovery of ``naked'' quasars with no host galaxies \citep[e.g.,][]{Bahcall1994}. However, these ``naked'' quasars were later found to be hosted by normal elliptical galaxies with improved smoothing of the HST images \citep{McLure1999,McIntosh1999}. Simulations have suggested another special class of SMBHs that may not follow the SMBH-host correlation: the ejected SMBHs \citep{Loeb2007,Haiman2009,Ricarte2021}. A SMBH binary in a gas-rich merger could be ejected as a SMBH remnant carrying an accretion disk. A SMBH remnant of an SMBH binary with similar masses could have a recoil speed of thousands of $\kms$. The ejected SMBH could transverse a considerable distance from the merged galaxy and be observed as an off-centered quasar if the ejected SMBH happens to pass through a dense molecular cloud. 

Red and obscured AGN are very important candidates for understanding the early quasar phases, galaxy quenching, and the enrichment of the intergalactic medium. 
Interactions between galaxies can help remove the angular momentum of the cold gas in the outskirts of galaxies. The inflowed gas can fuel the star formation in the galaxies, feed the central SMBH, and trigger AGN. The systems are then dusty, gaseous, and usually observed as red and obscured AGN. The radiation from AGN could in return power feedback and affect the evolution of their host galaxies. Strong outflows can clear up the gas reservoir in the host galaxies, shut down the star formation, enrich the environments, and make the AGN visible in optical. Outflows are widely used in understanding the co-evolution between SMBHs and their host galaxies, the lack of luminous quasars in the luminosity function, and the quenching of the star formation of galaxies. 

Spatially extended ionized gas and powerful outflow winds have been observed in a few high-redshift obscured quasars \citep{Cai2017,Fluetsch2021,Vayner2021,Lau2022}. These sources were pre-identified in large surveys and followed-up with hours of observation time for spatially resolved spectra on the Large Binocular Telescope/Medium-Dispersion Grating Spectroscopy, Keck/Keck Cosmic Web Imager, Very Large Telescope/Multi Unit Spectroscopic Explorer, etc.

In order to search for AGN with high EW, we need a spectroscopic survey that does not require continuum imaging pre-selection such as the Hobby-Eberly Telescope Dark Energy Experiment (HETDEX, \citealt{Gebhardt2021}). HETDEX is a spectroscopic survey with no photometric pre-selection (magnitude/color/morphology). All sources within the footprint of the survey are observed with a set of 78 Integral Field Units (IFUs) consisting of 34,944 fibers and a 18-minute exposure. HETDEX enables spectroscopic detection of the AGN hosted by galaxies that may be fainter than the detection limit of the corresponding photometric observations. Additionally, there is no need to perform follow-up observations for the extended sources on other instruments. The spatially resolved spectra can be obtained directly.

In this paper, we introduce an AGN (HETDEX J115031.93+504850.4, shortened to J1150+5048 in this paper) with extremely high EW ($\rm EW_{Ly\alpha+N\,V,rest} \gtrsim 921\,\AA$) at $z\sim2.24$ from the HETDEX survey. Section \ref{sec_data} briefly summarizes the first AGN catalog of the HETDEX survey. 
In Section \ref{sec_info}
, we present the basic information of J1150+5048, why it is selected for study, and the detailed spatially resolved properties of J1150+5048 with its narrow-band flux maps. We discuss the possible explanations for this high EW AGN in Section \ref{sec_discuss}. We summarize our discovery in Section \ref{sec_summary}. 

\section{The HETDEX AGN Catalog}
\label{sec_data}

HETDEX \citep{Gebhardt2021} is an ongoing spectroscopic survey (3500 \AA\ - 5500 \AA) on the upgraded 10-m Hobby-Eberly Telescope (HET, \citealt{Hill2021}). It uses the Visible Integral field Replicable Unit Spectrograph \citep[VIRUS;][]{Hill2021} to record spectra of every object falling within its field of view. A typical exposure contains 34,944 spectra, most of which capture ``blank'' sky. The primary goal of this survey is to measure the large-scale structure at $z\sim3$ using $\rm Ly\alpha$ emitters (LAEs) as tracers. The HETDEX survey is expected to be active from 2017 to 2024, and eventually will cover 540 deg$^2$ with a filling factor of 1 in 4.6.

The first AGN catalog of the HETDEX survey is presented in \cite{Liu2022} (Paper I\null). Here we briefly summarize the sample identification. AGN candidates are identified by requiring at least two significant AGN emission lines, such as the $\rm Ly\alpha$ and \ion{C}{4} $\lambda1549$ line pair, or with a single broad emission line with FWHM$>$1000~km\,s$^{-1}$, free of any pre-selection based on imaging (magnitude, morphology, or color). Each candidate AGN is then confirmed by visual inspection. This catalog contains 5,322 AGN, covering an effective sky coverage of 30.61 deg$^2$ and a redshift range of $0.25<z<4.32$. Measurements from the overlap regions with the Hyper Suprime-Cam (HSC) imager of the Subaru telescope from the HSC-HETDEX joint survey (HSC-DEX; $5\sigma$ depth is $r\sim25$ mag; S15A, S17A, S18A, PI: A. Schulze, and S19B, PI: S. Mukae) and the HSC Subaru Strategic Program (HSC-SSP; $5\sigma$ depth is $r\sim26$ mag ;\citealt{Aihara2019}) show that the median $r$-band magnitude of our AGN catalog is 21.6 mag, with 34\% of the objects having $r > 22.5$. Approximately 2.6\% of the HETDEX AGN are not detected at $>5\sigma$ confidence.

\section{J1150+5048}
\label{sec_info}
\begin{figure} 
\centering
\includegraphics[width=\textwidth]{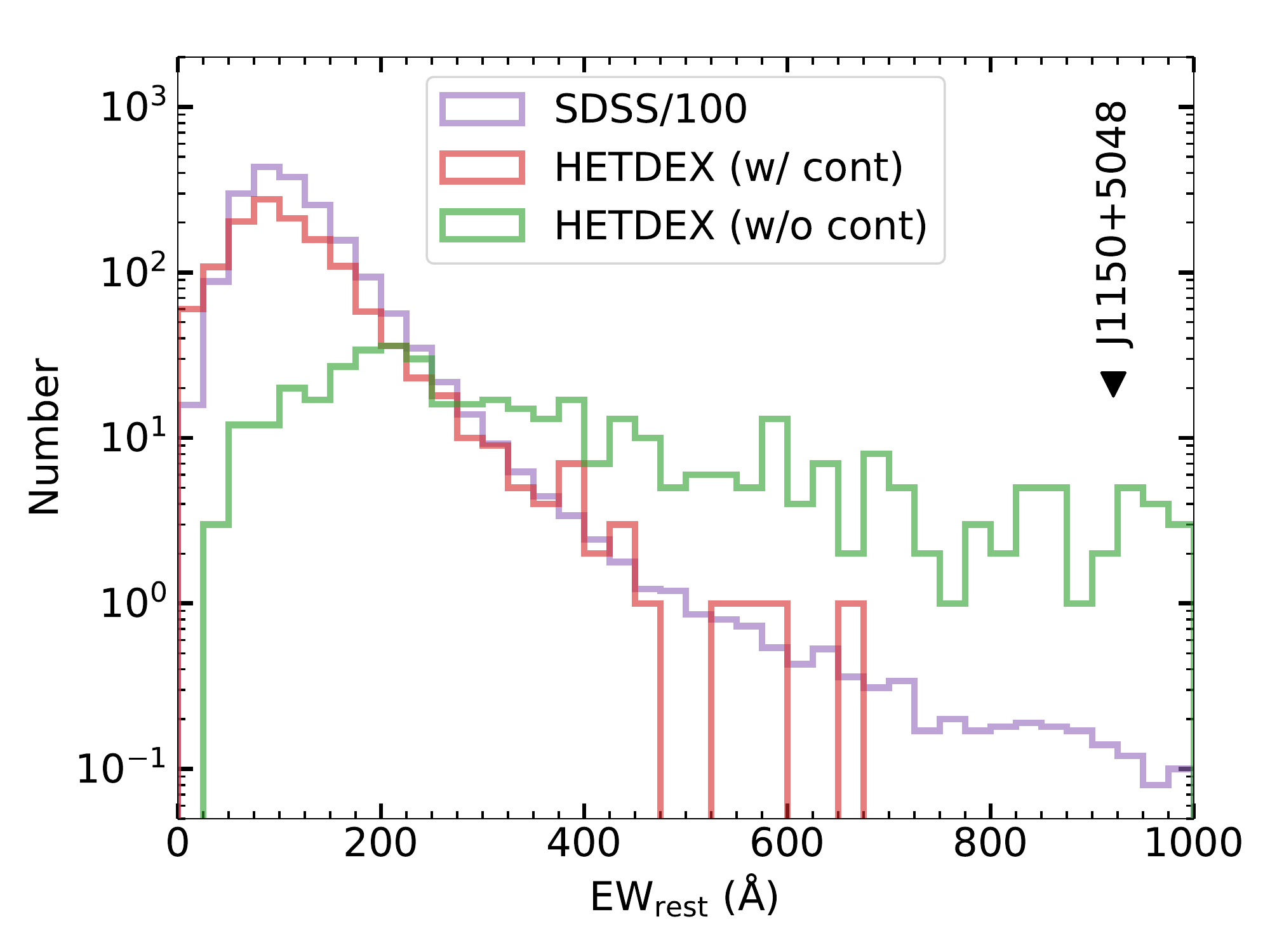}
\caption{Distributions of the rest-frame EW of the Ly$\alpha+$\ion{N}{5} $\lambda1241$ emission. HETDEX AGN with continuum detection ($>1\sigma$) are shown in the red histogram. HETDEX AGN without continuum detection ($<1\sigma$) are shown in the green histogram. SDSS quasars are indicated by purple histogram. The number of SDSS quasars in each bin is divided by 100 for presentation purposes. $\rm EW_{(Ly\alpha+N\,V),rest}$ of J1150+5048 is marked by the downward triangle.}
\label{f_ew}
\end{figure}

\begin{figure*} 
\centering
\includegraphics[width=\textwidth]{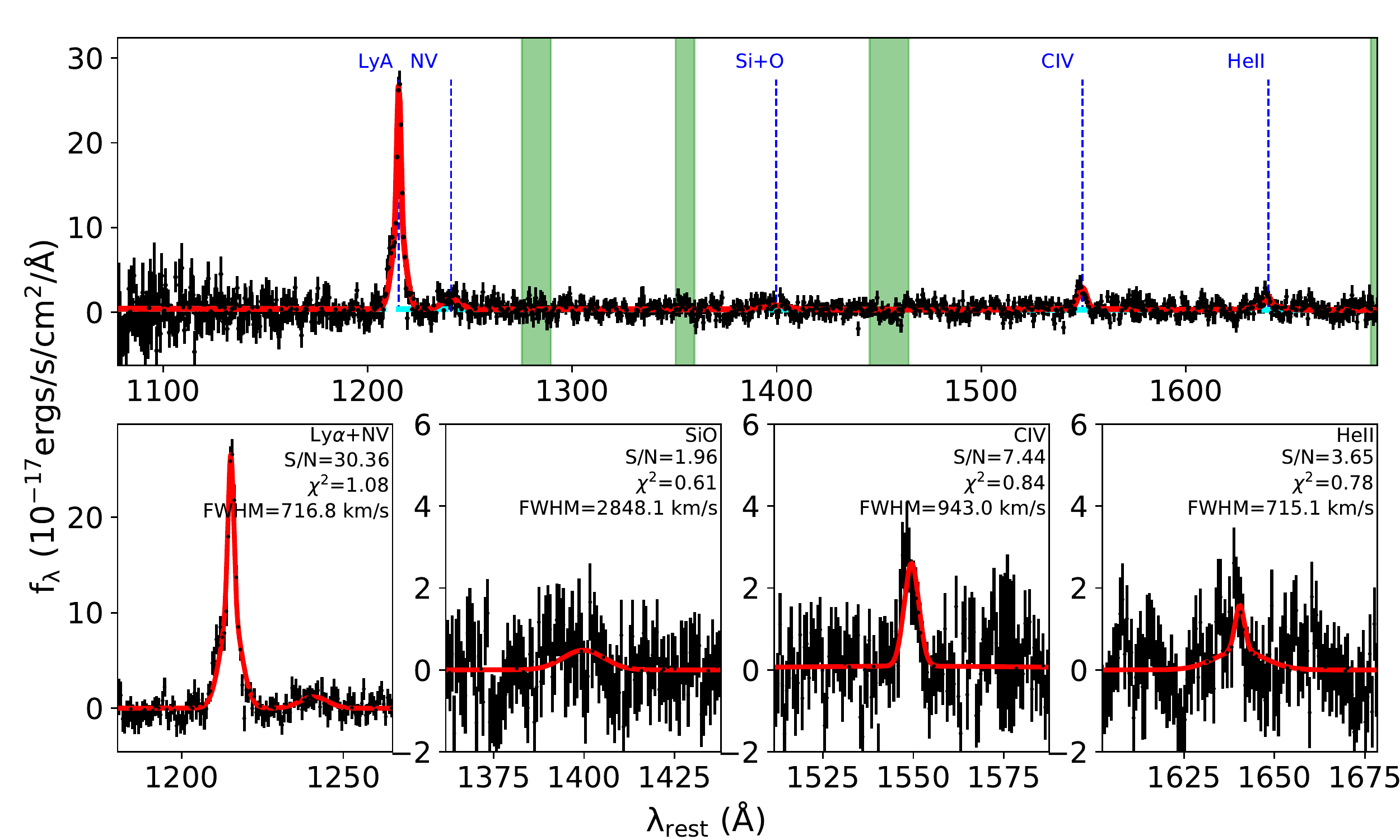}
\caption{The HETDEX spectrum of J1150+5048 at 
the HETDEX coordinate in Table \ref{t_info}.
Black data points with error bars are the observed spectrum shown in the rest-frame. The red line is our best fit to the spectrum. The green shaded areas indicate the continuum windows used in the continuum fit. The cyan dashed line is our best-fit continuum model. The four panels in the bottom row are the continuum subtracted local areas of the four emission lines.}
\label{f_spec}
\end{figure*}

\begin{table*} 
\centering
\setlength{\tabcolsep}{9pt}
\begin{tabular}{|c|c|c|c|c|c|}
\hline\hline
\multicolumn{6}{c}{Coordinate} \\ \hline
 \multicolumn{2}{|c|}{ (R.A., Dec.) (HETDEX)}      & \multicolumn{2}{c|}{ (R.A., Dec.) (WISE)}       & \multicolumn{2}{c|}{redshift}  \\ \hline 
 \multicolumn{2}{|c|}{ (177.633030, 50.813999)} & \multicolumn{2}{c|}{ (177.632964, 50.814293)}  & \multicolumn{2}{c|}{2.24}   \\ \hline
\hline
\multicolumn{6}{c}{Rest-frame equivalent width (\AA)} \\ \hline
$\rm EW_{Ly\alpha}$ & $\rm EW_{N\,V}$ & $\rm EW_{Ly\alpha+N\,V}$ & $\rm EW_{Si\,IV+O\,IV]}$ & $\rm EW_{C\,IV}$ & $\rm EW_{He\,II}$ \\ \hline
  $>$747            & $>$174          & $>$921                   & --              & $>$177         & $>$86           \\ \hline
\hline
\multicolumn{6}{c}{Broad-band Photometry} \\ \hline
 $g_{\rm AB}$ & $r_{\rm AB}$   & $\rm W1_{Vega}$ &  $\rm W2_{Vega}$ & $\rm W3_{Vega}$ & $\rm W4_{Vega}$ \\ \hline 
$>$24.23      & 24.57$\pm$0.13 & 17.50$\pm$0.14  &  16.82$\pm$0.29  & 12.47$\pm$0.38  & $>$8.93        \\ \hline
\end{tabular}
\caption{Basic information for J1150+5048}

\begin{tablenotes}[flushleft]
\scriptsize
\item 1) The Rest-frame EWs are all lower limits, because the continuum level in the [1275, 1290]\,\AA\ window at $(1.4\pm8)\times10^{-18}\, erg\,s^{-1}cm^{-2}\text{\AA}^{-1}$ used in the calculation of EWs is a non-detection.
\item 2) $g_{\text{AB}}$ is measured from the HETDEX spectrum (see Davis et al. in preparation for more details) at the HETDEX coordinate. The continuum of J1150+5048 is a non-detection in $g$-band. 24.23 is $1\sigma$ lower limit.
\item 3) $r_{\text{AB}}$ is measured from HSC-DEX at the WISE pointing. The $5\sigma$ depth of this field is $r=25.12$ (AB mag).
\item 4) W1-W4 are taken from the ALLWISE catalog \citep{Cutri2014}. W1, W2, W3, and W4 are the filters at 3.4, 4.6, 12, and 22 \micron, respectively. W4 is a non-detection.
\end{tablenotes}
\label{t_info}
\end{table*}

Figure \ref{f_ew} shows the distribution of the rest-frame EW of the Ly$\alpha+$\ion{N}{5} $\lambda1241$ emission of the HETDEX AGN with continuum detection ($>1\sigma$; red), that of the HETDEX AGN without continuum detection ($<1\sigma$; green), and that of the latest SDSS quasar catalog (purple; \citealt{Paris2018,Rakshit2020}). Similar with the SDSS quasars, the number of the HETDEX AGN detected with continuum decreases with $\rm EW_{(Ly\alpha+N\,V),rest}$. The green histogram shows the distribution of the lower limits of $\rm EW_{(Ly\alpha+N\,V),rest}$ for the HETDEX AGN not detected with continuum. In this paper, we study J1150+5048 as a representative case of the high EW AGN population with no continuum detection as it is detected with significant emission lines ($>7\sigma$) at Ly$\alpha+$\ion{N}{5} $\lambda1241$, \ion{C}{4} $\lambda1549$, and moderate emission line ($\sim4\sigma$) at \ion{He}{2} $\lambda1640$ (Figure \ref{f_spec}). Additionally, it has very deep $r$-band imaging from HSC-DEX. It is a type-II AGN with narrow lines ($\rm FWHM<1000\,\kms$). The fitting of the spectrum is detailed in Paper I. We fit a power-law continuum to the wavelength windows highlighted by the green shaded areas in Figure \ref{f_spec}. The continuum subtracted emission lines are then fit with two Gaussian profiles if there is a significant broad component, otherwise a single narrow Gaussian profile is fit for each emission line.

Table \ref{t_info} lists the basic information of J1150+5048. 
The EWs are measured directly from the ratio between the line flux and the best-fit continuum at the lines from the spectrum in Figure \ref{f_spec}. The continuum level is $<1\sigma$ in the spectrum, so the measured EWs are only lower limits. The lower limit of the rest-frame EW of the Ly$\alpha+$\ion{N}{5} $\lambda1241$ emission of J1150+5048 is 921\,\AA. This is significantly higher than the typical $\rm EW_{Ly\alpha+N\,V}$ at $\sim$100\,\AA\ as indicated by the $\rm EW_{Ly\alpha+N\,V}$ where the number of SDSS quasars is highest in Figure \ref{f_ew}.

\begin{figure*} 
\centering
\includegraphics[width=0.38\textwidth]{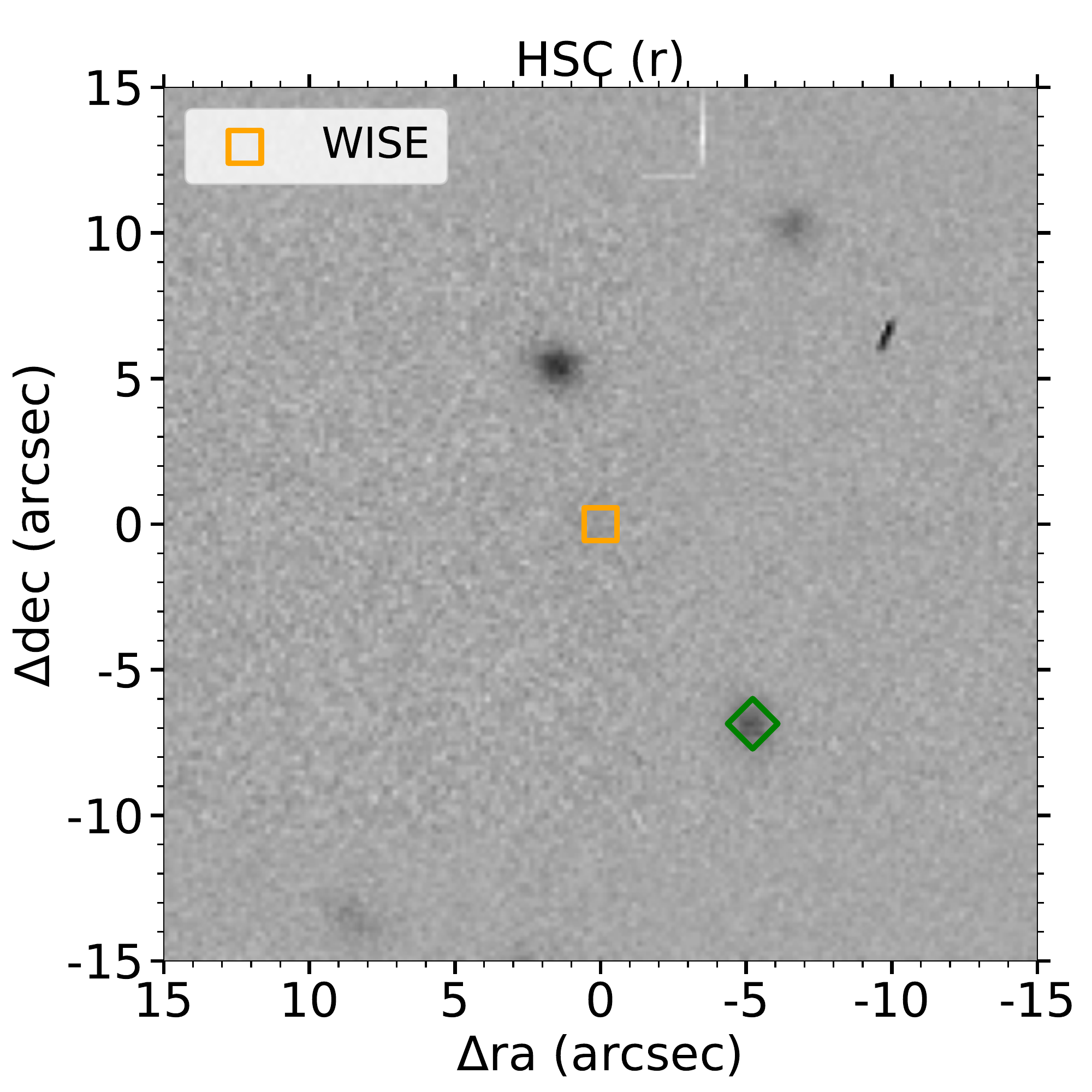}
\includegraphics[width=0.38\textwidth]{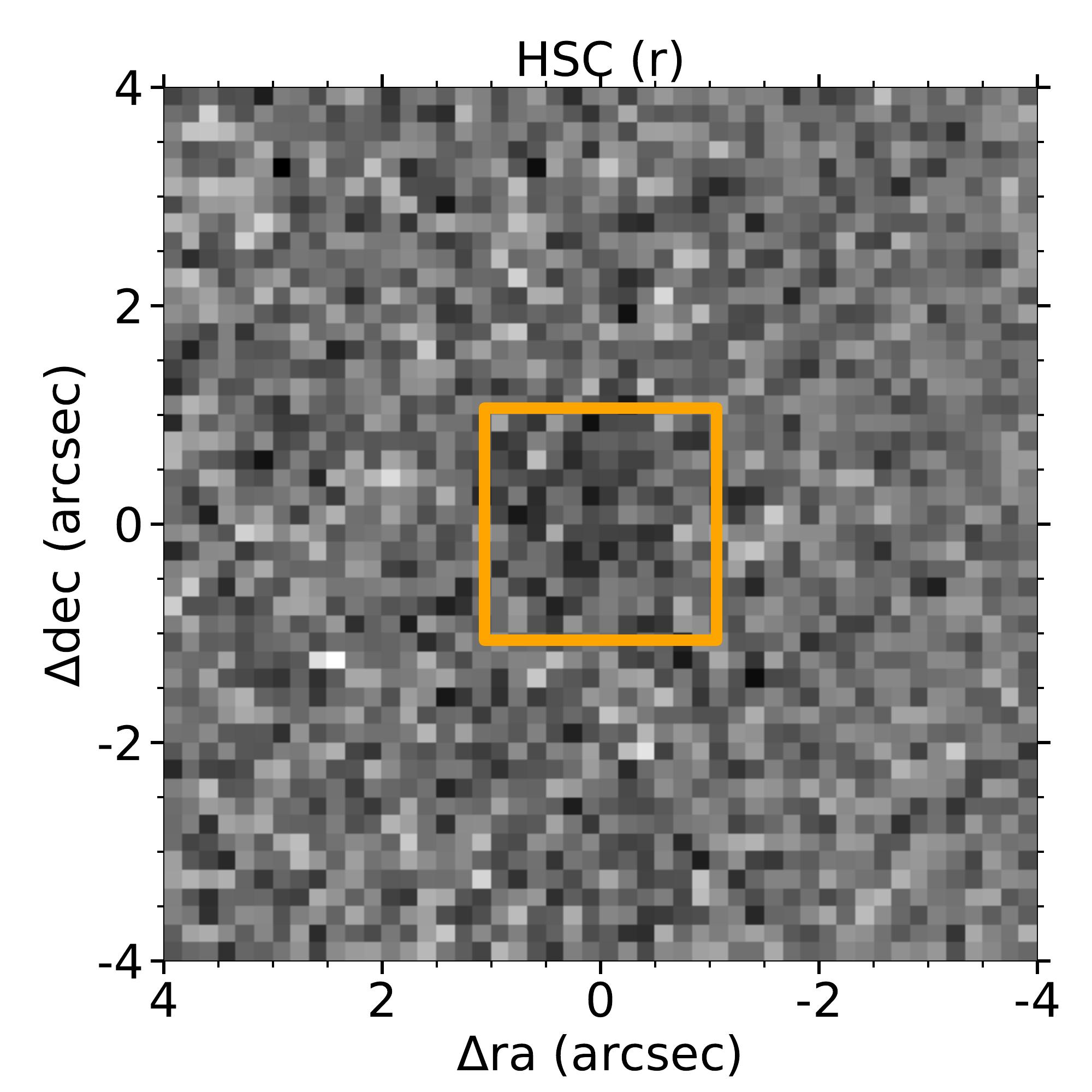}\\
\includegraphics[width=0.46\textwidth]{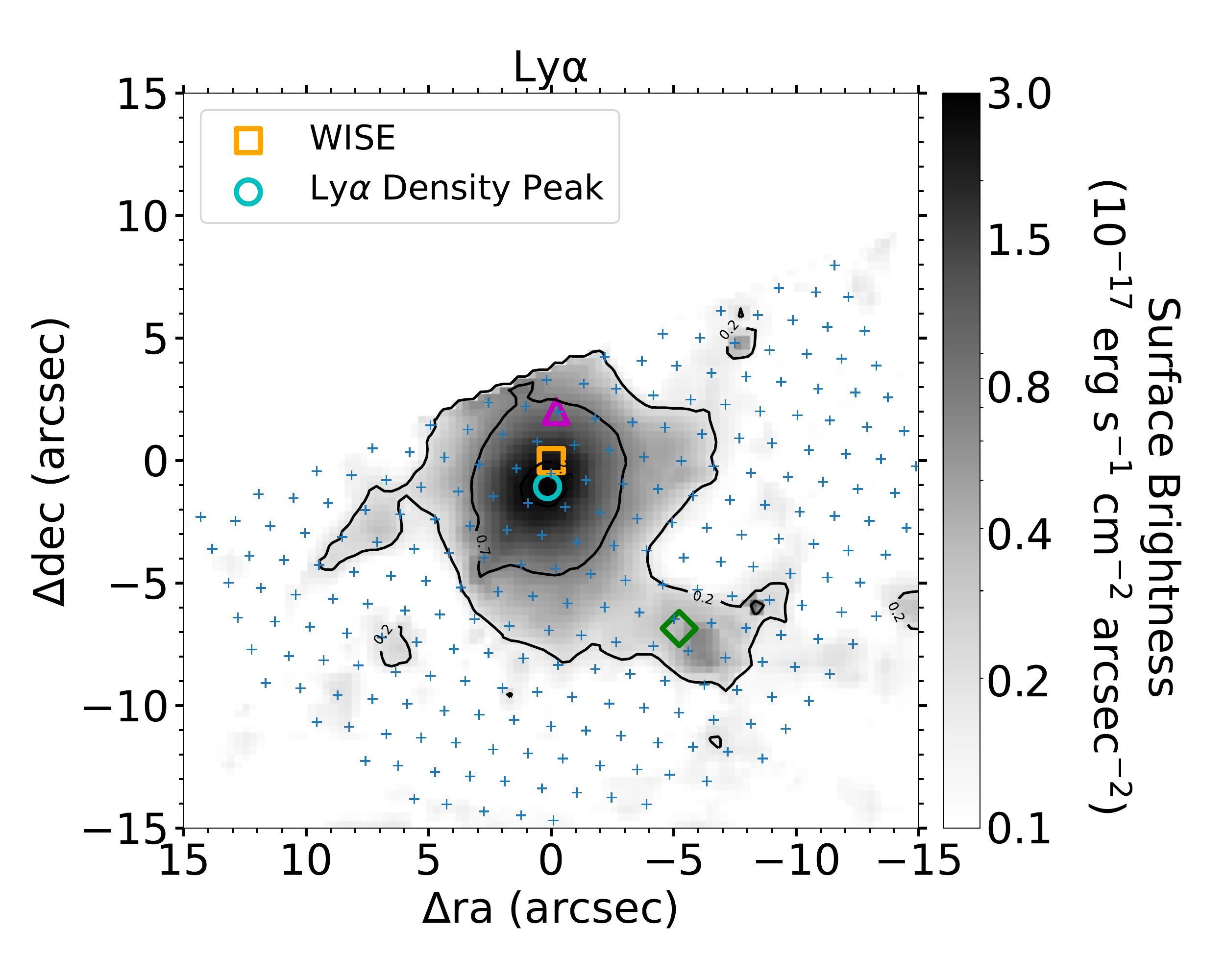}\\
\includegraphics[width=0.46\textwidth]{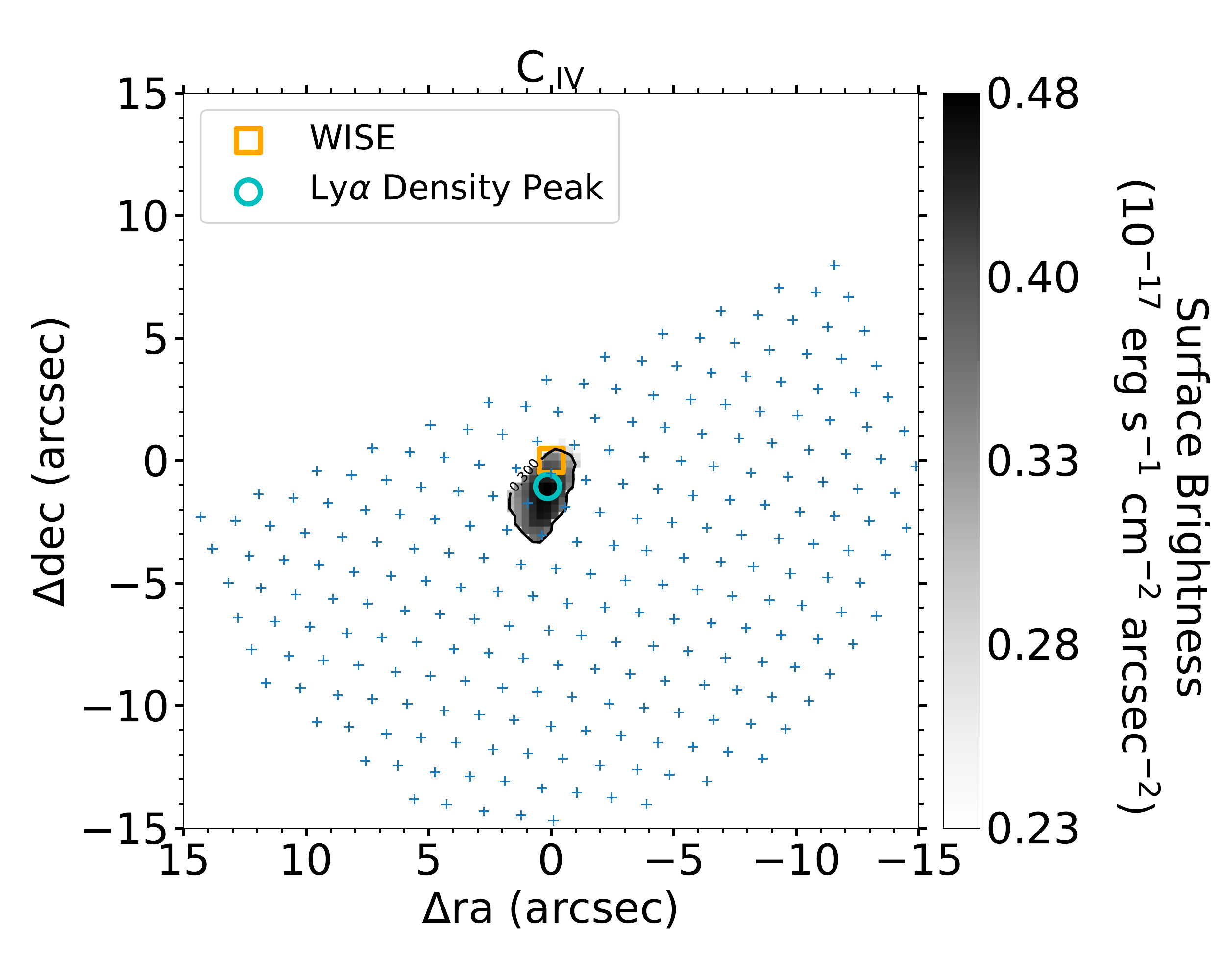}
\includegraphics[width=0.46\textwidth]{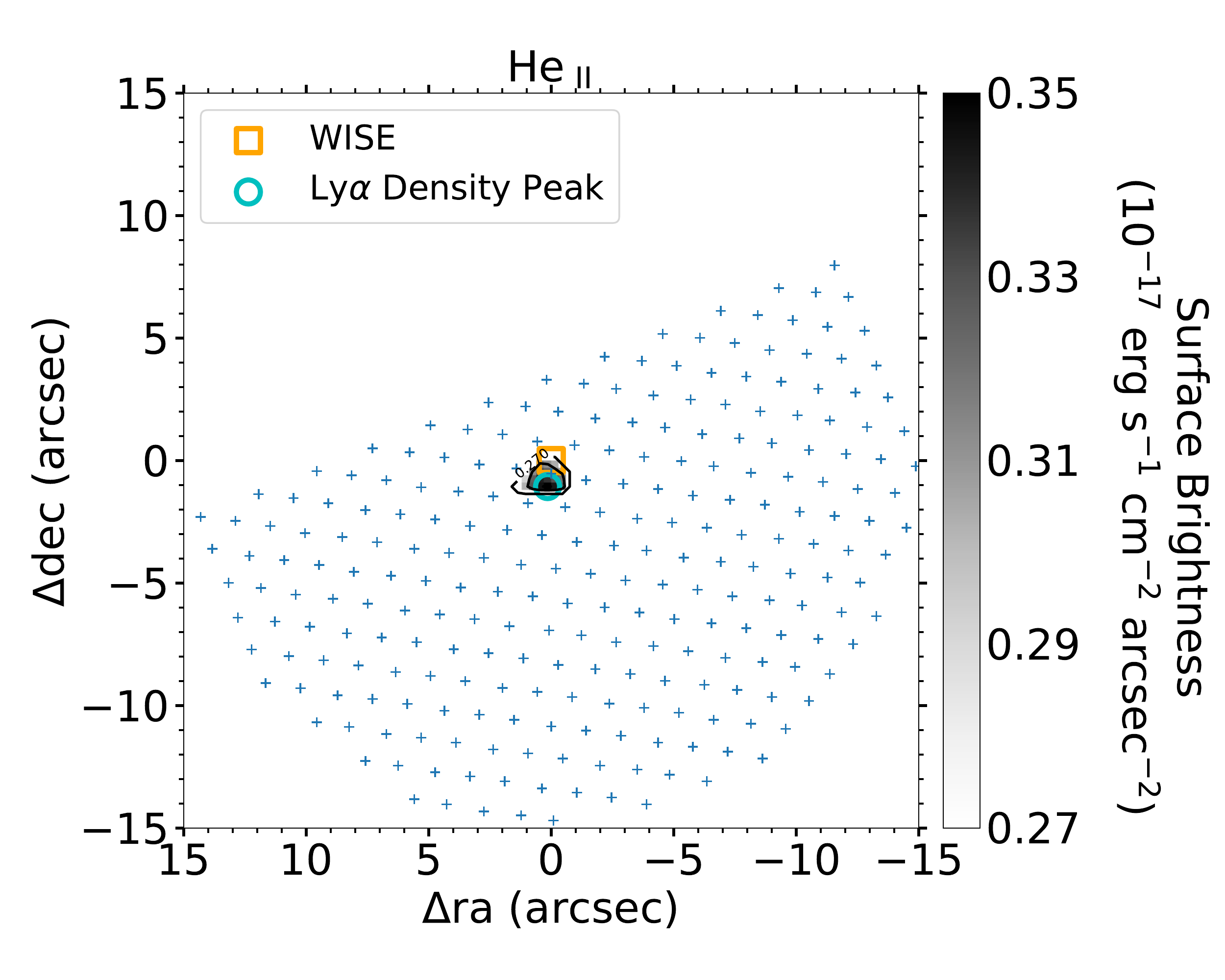}\\
\caption{Top panels: The $r$-band image and the expanded version from HSC-DEX. The pointing of the IR detection in the WISE survey is marked with an orange square. The green diamond marks another HSC-DEX $r$-band detection (22.5 mag) that is covered by the HETDEX fibers in this field. 
Middle panel: The narrow-band image of the $\rm Ly\alpha$ emission line derived from the HETDEX data. The small blue plus signs mark the positions of all HETDEX fibers in this field. The cyan circle marks the density peak of the Ly$\alpha$ line, which is also the recorded HETDEX coordinate in Table \ref{t_info}. The magenta triangle is a random position chosen to demonstrate the spatial changes of the Ly$\alpha$ line profiles in Figure \ref{f_BR}. Bottom panels: The narrow band images of the \ion{C}{4} $\lambda1549$ (left) and \ion{He}{2} $\lambda1640$ (right) emission lines. The seeing of the HETDEX observation is $1\farcs6$ (FWHM). All images are centered at the coordinate of WISE (continuum center) in Table \ref{t_info}.}
\label{f_image}
\end{figure*}

Figure \ref{f_image} displays the $r$-band cutout from HSC-DEX in the upper panels and the narrow band images ($\pm 20$\,\AA) of the $\rm Ly\alpha$, \ion{C}{4} $\lambda1549$, \ion{He}{2} $\lambda1640$ emission lines from HETDEX in the remaining three panels. Only spatial pixels with the signal-to-noise ratio of emission lines greater than 1 are used to generate the narrow band images.

\begin{figure*} 
\centering
\includegraphics[width=0.8\textwidth]{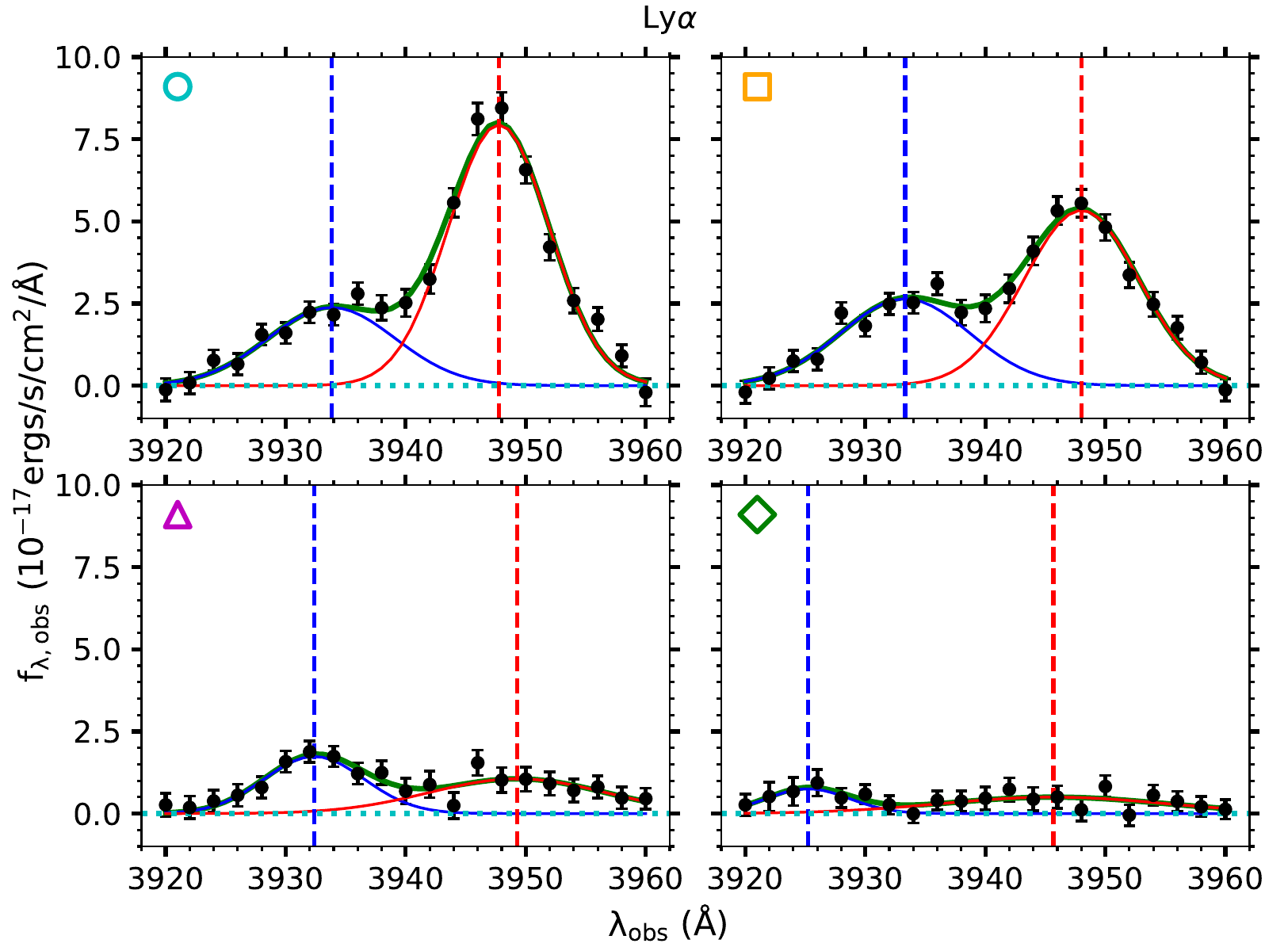}\\
\includegraphics[width=0.48\textwidth]{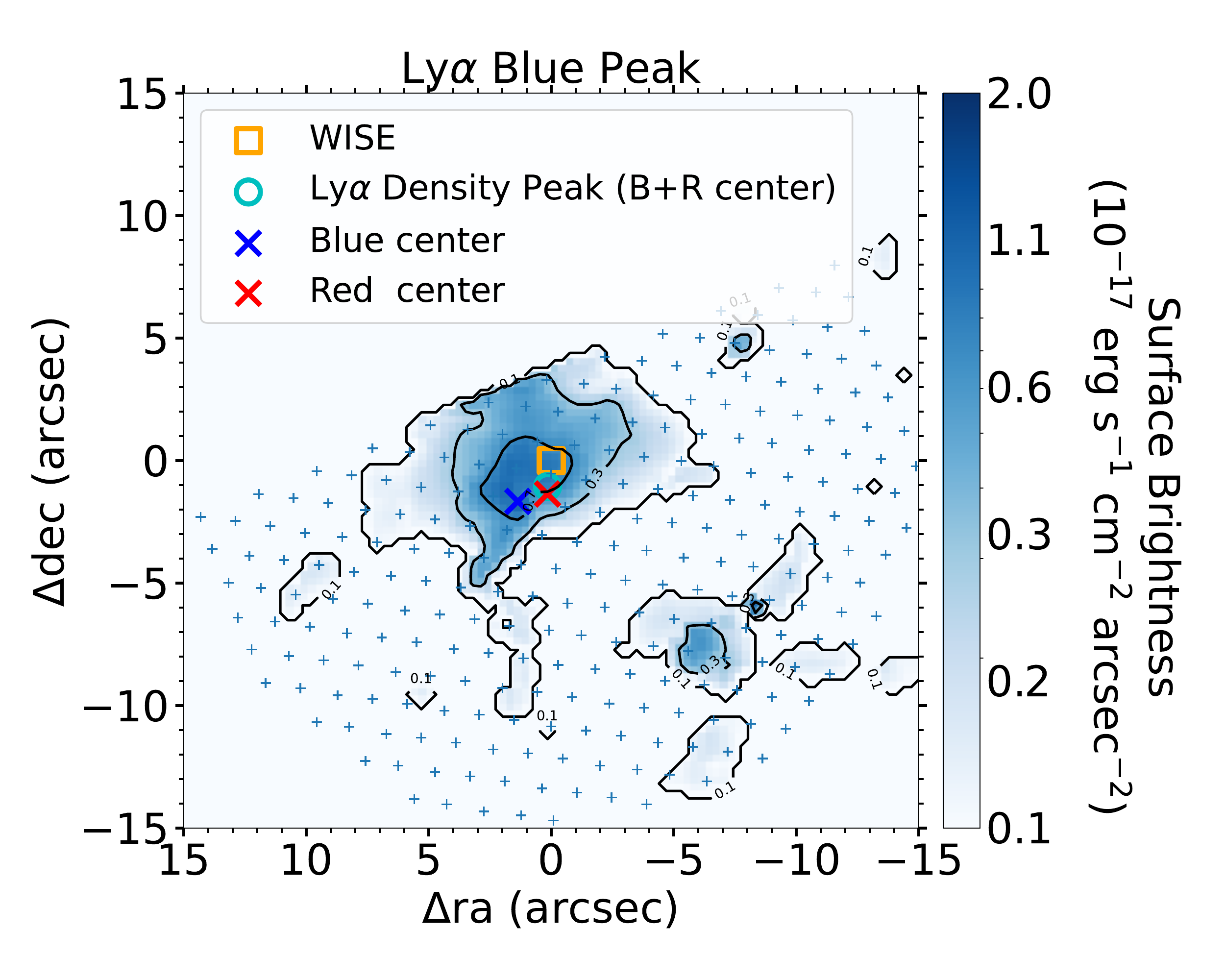}
\includegraphics[width=0.48\textwidth]{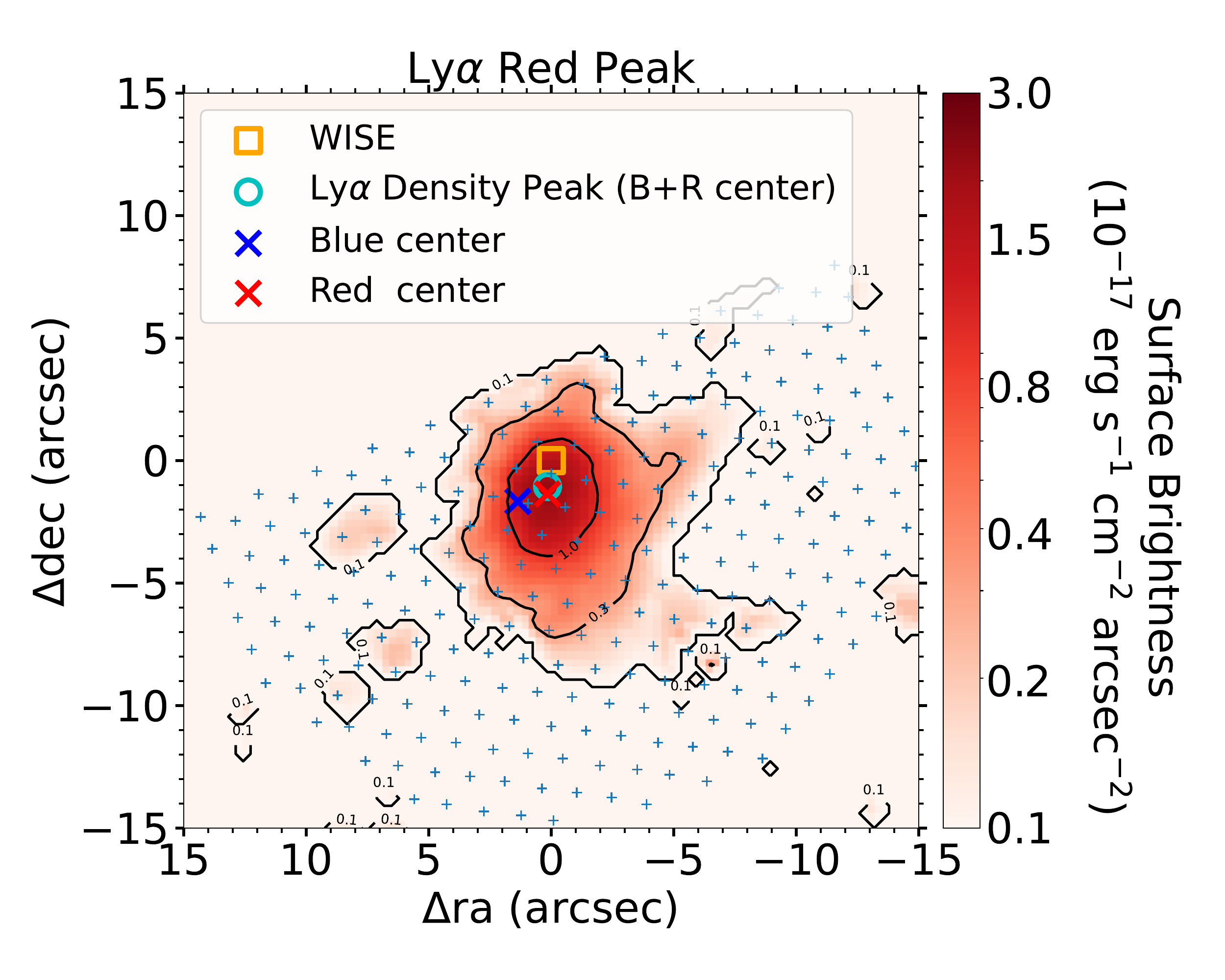}\\
\caption{Upper two rows: The spectra of Ly$\alpha$ at the four positions marked by the cyan circle (Ly$\alpha$ density peak), the orange square (the WISE pointing), the magenta triangle (a random position), and the green diamond (another $r$-band detection) in the narrow-band image of the Ly$\alpha$ emission line in Figure \ref{f_image}. Black data points with error bars are the observed spectra. Our best-fit double-Gaussian model is shown by the green curve. The blue and red solid curves are the best-fit blue component and red component. The blue and red vertical lines mark the best-fit central wavelengths of the two peaks. Bottom panels: The surface density map of the spectral decomposed blue peak and red peak of Ly$\alpha$. The orange square again marks the position of the WISE detection. The blue and the red crosses show the positions where the blue peak and the red peak have the highest line flux respectively.}
\label{f_BR}
\end{figure*}

The flux from the $\rm Ly\alpha$ emission line region is highly spatially resolved in J1150+5048 as shown by the middle panel of Figure \ref{f_image}. The seeing of this HETDEX observation is $1\farcs6$ (FWHM). The Ly$\alpha$ emission is clearly more extended than the other two emission lines. It spans a region of $\sim10\arcsec$ (85 kpc) in diameter.
The recorded R.A. and Dec. of HETDEX in Table \ref{t_info} is the coordinate where the $\rm Ly\alpha$ line flux is highest (the cyan circle in Figure \ref{f_image} and Figure \ref{f_BR}). The recorded R.A. and Dec. of WISE in Table \ref{t_info} is the coordinate of the continuum detection, taken from the WISE catalog \citep{Cutri2014}. All images in Figure \ref{f_image} and Figure \ref{f_BR} are centered at the WISE coordinate. The offset between the emission-line center (HETDEX coordinate) and the continuum center (WISE coordinate) is $1\farcs1$.

The emission lines of J1150+5048 are strongly bimodal with a red peak at $z\sim2.249$ and a blue peak at $z\sim2.236$ as shown in Figure \ref{f_BR}. The recorded redshift of $z\sim2.24$ in Table \ref{t_info} is a combination of the red peak and the blue peak.

J1150+5048 is a non-detection ($<1\sigma$) in $g$-band measured from the HETDEX spectrum at the HETDEX coordinate. It has an IR detection in the WISE catalog \citep{Cutri2014} with the W4 band (22 \micron) being a non-detection. When measuring at the pointing of the IR detection from HSC-DEX, its $r$-band magnitude is 24.57. The $\rm 5\sigma$ limiting magnitude of this field is $r=25.12$. 
The spatial distribution of the $r$-band flux of J1150+5048 (top panels in Figure \ref{f_image}) is strikingly different from the typical 24-25 mag sources (see the first two examples of Figure 12 in Paper I): It is too diffuse that it might be dominated by a spatially extended emission line, such as \ion{C}{3}] $\lambda 1909$, rather than the true continuum.

J1150+5048 is a non-detection in the Faint Images of the Radio Sky at Twenty-Centimeters survey (FIRST, \citealt{Becker1995}), while it has a radio detection at the location of the WISE source at $S_{\text{150\,MHz}} = 2\,m\text{Jy}$ from the LOw-Frequency ARray (LOFAR) Two-metre Sky Survey (LoTSS) \citep{Shimwell2022}.


Figure \ref{f_BR} presents the $\rm Ly\alpha$ line profiles at the four representative positions marked in the Ly$\alpha$ narrow band image of Figure \ref{f_image} in the upper two rows. The $\rm Ly\alpha$ emission line clearly has two distinctive peaks. We decompose the $\rm Ly\alpha$ emission line into a blue peak and a red peak with a double-Gaussian model. The velocity offset between the two peaks ($\sim 1100\,\kms$) does not change significantly with location. The bottom two panels shows the surface density maps of the decomposed blue peak and red peak respectively. The blue cross and the red cross mark the positions where the flux is highest for the blue peak and the red peak. The separation between the two centers is $\sim1\farcs2$ (10.1 kpc). For most of the spatial pixels, the blue peak is weaker than the red peak. This asymmetry is an evidence of outflows, as the near side of $\rm Ly\alpha$ is resonantly scattered by an optically thick medium. The \ion{C}{4} $\lambda1549$ and \ion{He}{2} $\lambda1640$ emission lines also display double peak profiles; however, the two lines are not sufficiently strong for spatial decomposition as was done for Ly$\alpha$.

\begin{figure*} 
\centering
\includegraphics[width=0.48\textwidth]{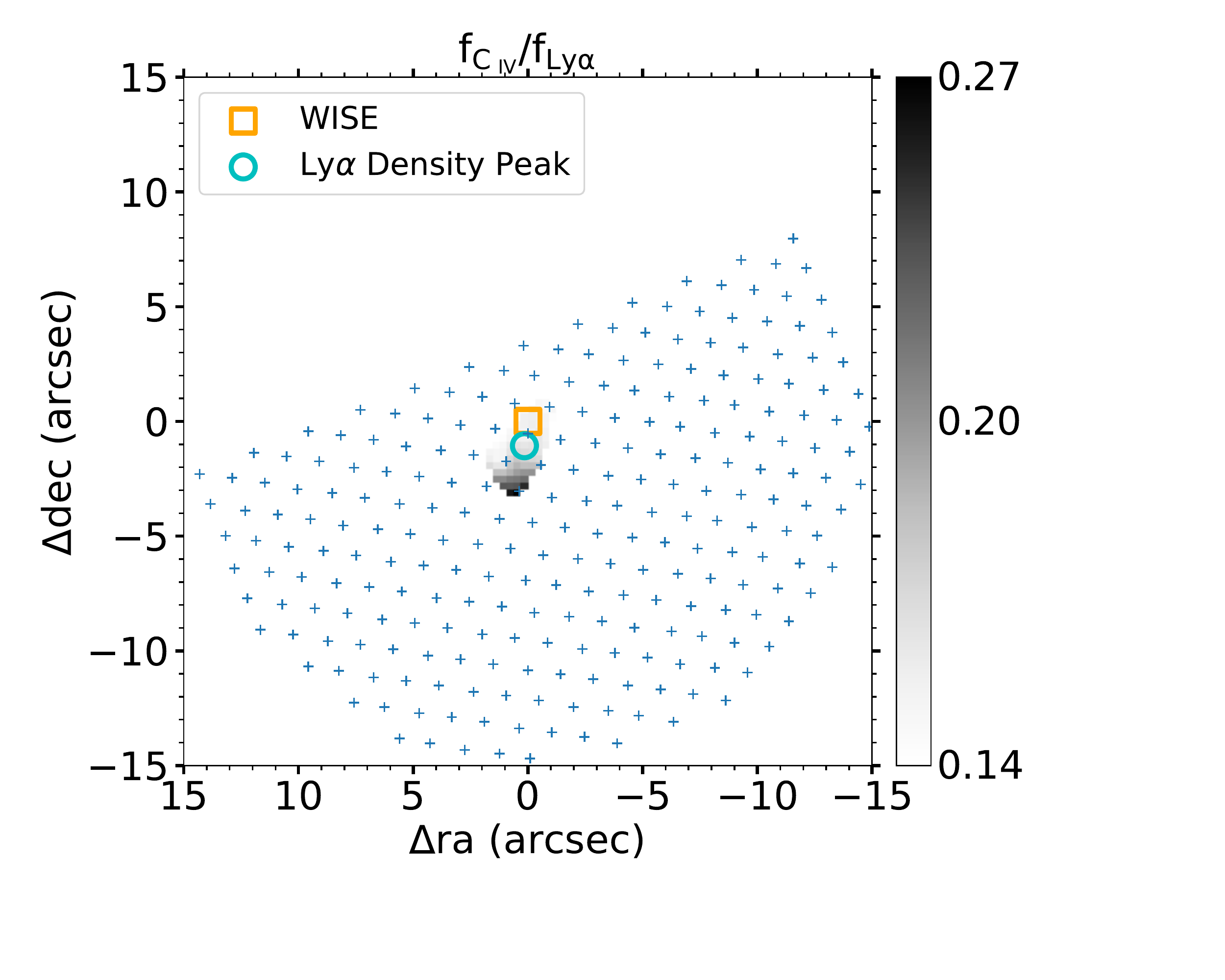}
\includegraphics[width=0.48\textwidth]{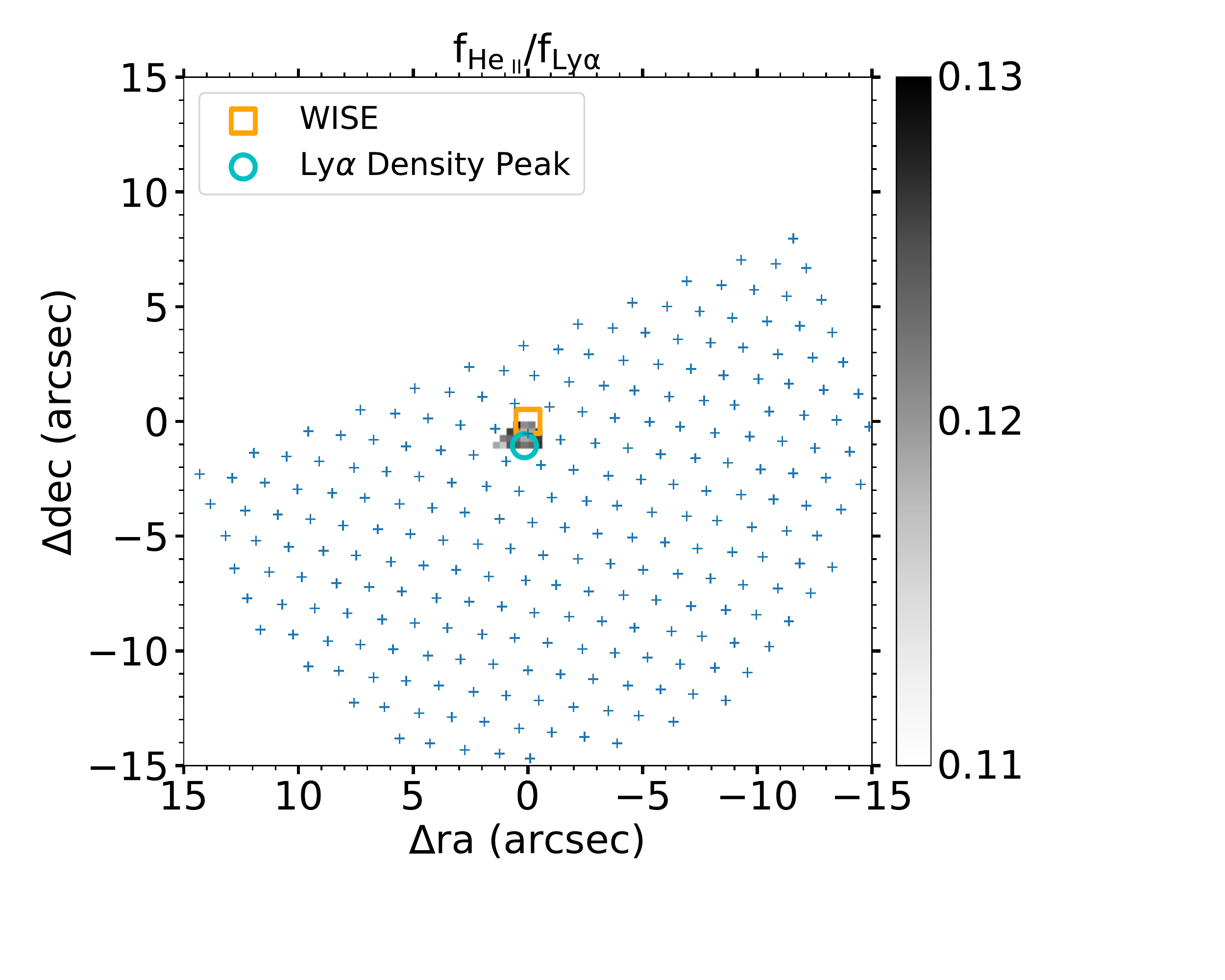}\\
\caption{Left: The line flux ratio map of the \ion{C}{4} $\lambda1549$ emission over the $\rm Ly\alpha$ emission. Only spatial pixels with 
both emission lines detected at $>1\sigma$ level are used in the map.
Labels are the same with the ones in the narrow-band images in Figure \ref{f_image}. Right: Similar plot with the left panel, but for the line ratio of the \ion{He}{2} $\lambda1640$ emission over the $\rm Ly\alpha$ emission.}
\label{f_ratio}
\end{figure*}

Figure \ref{f_ratio} shows the line flux ratio of the \ion{C}{4} $\lambda1549$ emission over the $\rm Ly\alpha$ emission in the left panel and that of the \ion{He}{2} $\lambda1640$ emission over the $\rm Ly\alpha$ emission in the right panel. The line ratio of \ion{C}{4} $\lambda1549/\rm Ly\alpha$ ranges from 0.1 to 0.3. The line ratio of \ion{He}{2} $\lambda1640/\rm Ly\alpha$ ranges from 0.11 to 0.13. The combination of the \ion{C}{4} $\lambda1549/$\ion{He}{2} $\lambda1640$ ratio and the \ion{C}{3}] $\lambda1909/$\ion{C}{4} $\lambda1549$ would provide diagnostics for the ionization levels and the metallicities given quasar photoionization models \citep[e.g.][]{Guo2020}. Unfortunately, the \ion{C}{3}] $\lambda1909$ emission is out of the HETDEX wavelength range. \cite{Lau2022} suggested that the the ionization parameter can be assumed to be $\lg U\sim-1$. The \ion{C}{4} $\lambda1549/$\ion{He}{2} $\lambda1640$ ratio of $\sim1$ in at the $\rm Ly\alpha$ density peak (shown by the cyan circle) would suggest a metallicity of $\rm\sim Z_{\sun}$. The spatially-integrated line ratio of $\sim3$ out to $\sim3\arcsec$ ($\sim$ 20\,kpc) corresponds to a metallicity of $\rm\sim 0.5\,Z_{\sun}$. 
The metallicity enrichment suggests centrally-driven outflows to the host galaxy.

\section{Discussion}
\label{sec_discuss}

There are several possibilities that can produce the high EW of J1150+5048; we discuss three possible explanations in this section.

\subsection{Collapsing Protogiant Elliptical Galaxy}
\label{sec_collap}

\cite{Adams2009} studied the famous radio loud AGN (B2 0902+34) at $z=3.4$ with VIRUS-prototype on the 2.7-m Harlan J. Smith Telescope. B2 0902+34 also has bimodal Ly$\alpha$ emission line profile, and the Ly$\alpha$ emission is extended with a radius of $\sim$ 50 kpc. The observed data was successfully reproduced with a model of a collapsing protogiant elliptical galaxy ($\rm \ge10^{12}\ M_\sun$). $z\sim2$ is a common era of galaxy formation. Massive galaxies and their formation have many signatures, most of which include a lot of emission in radio and infrared. B2 0902+34 is detected at 300 $m$Jy in the FIRST survey. J1150+5048 is also covered by FIRST, but there is no radio excess within $\pm 30\arcsec$ of the AGN. The radio detection of J1150+5048 in LoTSS at $S_{\text{150\,MHz}} = 2\,m\text{Jy}$ is also too weak compared to B2 0902+34. These may rule out the possibility of collapsing protogiant elliptical galaxy.

\subsection{Off-centered SMBH}
\label{sec_naked}

SMBHs can be ejected after merger event \citep{Loeb2007,Haiman2009,Ricarte2021}. If they happen to fall in a gas rich environment, the ejected SMBH can irradiate the inter-galactic medium (IGM) in its vicinity and appear as a naked BH with no host galaxy. The narrow band emission line flux maps in Figure \ref{f_image} show that J1150+5048 might be related with the other $r$-band detection covered by the HETDEX fibers marked by the green diamond which is $r_{\text{AB}}=22.5\,\text{mag}$. It has weak emission lines at similar redshift with J1150+5048 with a velocity offset of $\sim -600\,\kms$ as shown by the spectra in Figure \ref{f_BR}. If the green diamond is the original host and the WISE detection is the SMBH, then the separation between the two is $8\farcs6$ (72 kpc).

\subsection{Extremely Red Quasar with Strong Winds}
\label{sec_ulirg}

The optical broad band imaging could be heavily obscured by dust in the outskirts of the host galaxy. 
Extremely red quasars (ERQs) are first were first identified in \cite{Ross2015} with $r_{\text{AB}}-\text{W4}_{\text{Vega}}>14\,\text{mag}$, i.e. $r_{\text{AB}}-\text{W4}_{\text{AB}}>7.38\,\text{mag}$. They are revisited by \cite{Hamann2017} with the definition of $i_{\text{AB}}-\text{W3}_{\text{AB}}\ge4.6\,\text{mag}$ and the rest-frame $\rm EW_{CIV}\ge100\,\AA$. Although J1150+5048 is a non-detection in the W4 band and it lacks the observation in $i$-band, the color of $r_{\text{AB}}-\text{W3}_{\text{AB}}=6.93\,\text{mag}$ and high $\rm EW_{CIV}>177\,\AA$ indicates that J1150+5048 probably belongs to the ERQ population. 
The current photometric data is not sufficient to break the degeneracy among various models in the spectral energetic distribution. Photometric observations in more bands, such as $K$-band and $z$-band, might further help confirm whether J1150+5048 is an ERQ.

Many of ERQs are found in large-scale overdensities, we checked the full emission line catalog of the HETDEX survey (Cooper et al. in preparation) and only found one weak $\rm Ly\alpha$ emitter candidate within $\pm5\arcmin$ (2.5\,Mpc) at $z=2.24\pm0.03$ down to the detection limit of HETDEX at $g\sim24.5$ mag. J1150+5048 might not be in an large-scale overdensity.

Besides J1150+5048, we found that other high EW AGN in our catalog are all narrow-line AGN with $\rm FWHM\sim1000\,\kms$, corresponding to the velocity dispersion of $\sim400\,\kms$. \cite{Lau2022} also found that the $\rm Ly\alpha$ halo of the ERQ they studied is kinematically quiet with the velocity dispersion of $\sim300\,\kms$.

A significant fraction of our AGN sample has extended emission-line regions. 
\cite{Ouchi2020} collected emitters with measurements of their diffuse $\rm Ly\alpha$ emission and found that the radius of diffuse $\rm Ly\alpha$ emitters are correlated with the $\rm Ly\alpha$ luminosities. The $\rm Ly\alpha$ luminosity of their sample ranges from $\rm\sim10^{42}\,erg\,s^{-1}$ to $\rm\sim10^{45}\,erg\,s^{-1}$.
The spatially integrated $\rm Ly\alpha$ luminosity of J1150+5048 is $\rm10^{43.4}\,erg\,s^{-1}$. The $\rm Ly\alpha$ emission of J1150+5048 extends to $r\sim85$ kpc. J1150+5048 lies well within the scattered region of the correlation between the radius and the $\rm Ly\alpha$ luminosity in Figure 13 of \cite{Ouchi2020}.

The extended emission-line region can be explained either by central-driven outflows or by inflows from the circumgalactic medium. The blue peak of the $\rm Ly\alpha$ emission is always weaker than the red peak. This suggests that the near side is more heavily scattered than the far side, and the gas flows are outflows rather than inflows.
If the direction of the outflows are perpendicular to the line-of-sight direction, the separation between the blue peak and the red peak ($\sim 1100\,\kms$) would not change significantly with positions as was found in Section \ref{sec_info}. 

We estimate the outflow mass $M_{\text{outflow}}$ and the outflow rate $\dot{M}_{\text{outflow}}$ of J1150+5048 following Equation 5 and 6 in \cite{Fluetsch2021}. However, the H$\rm\alpha$ emission is not covered by the wavelength range of the HETDEX spectrum. We therefore make a simple assumption that the luminosity of the $\rm Ly\alpha$ emission is around 1.5 times that of the H$\rm\alpha$ emission following \citep{Allen1982}. The electron density $n_e$ should be carefully calculated from doublets such as the [\ion{O}{2}] $\lambda3726$/[\ion{O}{2}] $\lambda3729$ ratio and the [\ion{S}{2}] $\lambda6717$/[\ion{S}{2}] $\lambda6731$ ratio, but these emission lines are again out of the wavelength coverage of HETDEX. We then take the typical $n_e$ of outflows $\sim500\,cm^{-3}$ from \cite{Fluetsch2021}. With these two assumptions, the outflow mass and the outflow rate of J1150+5048 are then $\lg(M_{\text{outflow}}/\text{M}_{\sun})\sim8$ and 1.5 $\rm M_{\sun}\,yr^{-1}$.

As shown in Figure \ref{f_ew}, there are many other high-EW AGN in our catalog, although some are missing $r$-band imaging, and some do not have significant emission line detected besides $\rm Ly\alpha$. By the time the HETDEX survey is complete, we expect the final AGN sample to be about five times larger than the current one. We will systematically study all AGN with high EWs. It is expected that many of these high-EW AGN are similar to J1150+5048 with red colors and bipolar outflows. It is interesting to study the outflow properties, such as the outflow mass $M_{\text{outflow}}$ and the outflow rate $\dot{M}_{\text{outflow}}$, as a function of luminosity, reddening, and redshift.

\section{Summary}
\label{sec_summary}

We have identified an AGN (J1150+5048) with extremely high EW at $z\sim2.24$, with strong Ly$\alpha$, \ion{C}{4} $\lambda1549$, and \ion{He}{2} $\lambda1640$ emission lines and non-detected continuum in the HETDEX spectrum. The measured $\rm EW_{Ly\alpha+N\,V,rest}=921\,\AA$ is a lower limit of its rest-frame line strength at Ly$\alpha$. Extended emission is measured at $r=24.57$ in the deep $r$-band image ($r_{5\sigma}=25.12$) from the HSC-DEX survey. It has an IR detection in the WISE catalog, but it is a non-detection in the W4 band. The Ly$\alpha$ emission line is significantly extended in the narrow band image spanning $\sim$ 10\arcsec\ in diameter in the observation. The line profile of Ly$\alpha$ is strongly bimodal. The decomposed blue peak and red peak separated from each other with $1\farcs2$. The line-of-sight velocity offset between the two peaks is $\sim 1100 \kms$. 

Further statistical studies on the full high-EW AGN sample are needed to understand the relation among this sample, the ERQs, type-II AGN, and the AGN with diffuse ionized gas. These would provide key information in understanding the early quasar phases, galaxy quenching, and the enrichment of the intergalactic medium. The HETDEX survey is very efficient in such studies because the spatial resolved information is observed simultaneously while the emission-line sources are identified in the 18-min spectroscopic exposures with no pre-selection based on imaging.

\acknowledgments

HETDEX is led by the University of Texas at Austin McDonald Observatory and Department of Astronomy with participation from the Ludwig-Maximilians-Universit\"at M\"unchen, Max-Planck-Institut f\"ur Extraterrestrische Physik (MPE), Leibniz-Institut f\"ur Astrophysik Potsdam (AIP), Texas A\&M University, The Pennsylvania State University, Institut f\"ur Astrophysik G\"ottingen, The University of Oxford, Max-Planck-Institut f\"ur Astrophysik (MPA), The University of Tokyo, and Missouri University of Science and Technology. In addition to Institutional support, HETDEX is funded by the National Science Foundation (grant AST-0926815), the State of Texas, the US Air Force (AFRL FA9451-04-2-0355), and generous support from private individuals and foundations.

The Hobby-Eberly Telescope (HET) is a joint project of the University of Texas at Austin, the Pennsylvania State University, Ludwig-Maximilians-Universit\"at M\"unchen, and Georg-August-Universit\"at G\"ottingen. The HET is named in honor of its principal benefactors, William P. Hobby and Robert E. Eberly.

The authors acknowledge the Texas Advanced Computing Center (TACC) at The University of Texas at Austin for providing high performance computing, visualization, and storage resources that have contributed to the research results reported within this paper. URL: http://www.tacc.utexas.edu

\bibliography{highEW}

\begin{thebibliography}{}
\expandafter\ifx\csname natexlab\endcsname\relax\def\natexlab#1{#1}\fi

\bibitem[{{Adams} {et~al.}(2009){Adams}, {Hill}, \& {MacQueen}}]{Adams2009}
{Adams}, J.~J., {Hill}, G.~J., \& {MacQueen}, P.~J. 2009, \apj, 694, 314

\bibitem[{{Aihara} {et~al.}(2019){Aihara}, {AlSayyad}, {Ando}, {Armstrong},
  {Bosch}, {Egami}, {Furusawa}, {Furusawa}, {Goulding}, {Harikane}, {Hikage},
  {Ho}, {Hsieh}, {Huang}, {Ikeda}, {Imanishi}, {Ito}, {Iwata}, {Jaelani},
  {Kakuma}, {Kawana}, {Kikuta}, {Kobayashi}, {Koike}, {Komiyama}, {Li},
  {Liang}, {Lin}, {Luo}, {Lupton}, {Lust}, {MacArthur}, {Matsuoka}, {Mineo},
  {Miyatake}, {Miyazaki}, {More}, {Murata}, {Namiki}, {Nishizawa}, {Oguri},
  {Okabe}, {Okamoto}, {Okura}, {Ono}, {Onodera}, {Onoue}, {Osato}, {Ouchi},
  {Shibuya}, {Strauss}, {Sugiyama}, {Suto}, {Takada}, {Takagi}, {Takata},
  {Takita}, {Tanaka}, {Terai}, {Toba}, {Uchiyama}, {Utsumi}, {Wang}, {Wang}, \&
  {Yamada}}]{Aihara2019}
{Aihara}, H., {AlSayyad}, Y., {Ando}, M., {et~al.} 2019, \pasj, 71, 114

\bibitem[{{Allen} {et~al.}(1982){Allen}, {Barton}, {Gillingham}, \&
  {Carswell}}]{Allen1982}
{Allen}, D.~A., {Barton}, J.~R., {Gillingham}, P.~R., \& {Carswell}, R.~F.
  1982, \mnras, 200, 271

\bibitem[{{Bahcall} {et~al.}(1994){Bahcall}, {Kirhakos}, \&
  {Schneider}}]{Bahcall1994}
{Bahcall}, J.~N., {Kirhakos}, S., \& {Schneider}, D.~P. 1994, \apjl, 435, L11

\bibitem[{{Becker} {et~al.}(1995){Becker}, {White}, \& {Helfand}}]{Becker1995}
{Becker}, R.~H., {White}, R.~L., \& {Helfand}, D.~J. 1995, \apj, 450, 559

\bibitem[{{Cai} {et~al.}(2017){Cai}, {Fan}, {Yang}, {Bian}, {Prochaska},
  {Zabludoff}, {McGreer}, {Zheng}, {Green}, {Cantalupo}, {Frye}, {Hamden},
  {Jiang}, {Kashikawa}, \& {Wang}}]{Cai2017}
{Cai}, Z., {Fan}, X., {Yang}, Y., {et~al.} 2017, \apj, 837, 71

\bibitem[{{Cutri} {et~al.}(2021){Cutri}, {Wright}, {Conrow}, {Fowler},
  {Eisenhardt}, {Grillmair}, {Kirkpatrick}, {Masci}, {McCallon}, {Wheelock},
  {Fajardo-Acosta}, {Yan}, {Benford}, {Harbut}, {Jarrett}, {Lake}, {Leisawitz},
  {Ressler}, {Stanford}, {Tsai}, {Liu}, {Helou}, {Mainzer}, {Gettngs},
  {Gonzalez}, {Hoffman}, {Marsh}, {Padgett}, {Skrutskie}, {Beck}, {Papin}, \&
  {Wittman}}]{Cutri2014}
{Cutri}, R.~M., {Wright}, E.~L., {Conrow}, T., {et~al.} 2021, VizieR Online
  Data Catalog, II/328

\bibitem[{{Fluetsch} {et~al.}(2021){Fluetsch}, {Maiolino}, {Carniani},
  {Arribas}, {Belfiore}, {Bellocchi}, {Cazzoli}, {Cicone}, {Cresci}, {Fabian},
  {Gallagher}, {Ishibashi}, {Mannucci}, {Marconi}, {Perna}, {Sturm}, \&
  {Venturi}}]{Fluetsch2021}
{Fluetsch}, A., {Maiolino}, R., {Carniani}, S., {et~al.} 2021, \mnras, 505,
  5753

\bibitem[{{Gebhardt} {et~al.}(2021){Gebhardt}, {Mentuch Cooper}, {Ciardullo},
  {Acquaviva}, {Bender}, {Bowman}, {Castanheira}, {Dalton}, {Davis}, {de Jong},
  {DePoy}, {Devarakonda}, {Dongsheng}, {Drory}, {Fabricius}, {Farrow},
  {Feldmeier}, {Finkelstein}, {Froning}, {Gawiser}, {Gronwall}, {Herold},
  {Hill}, {Hopp}, {House}, {Janowiecki}, {Jarvis}, {Jeong}, {Jogee}, {Kakuma},
  {Kelz}, {Kollatschny}, {Komatsu}, {Krumpe}, {Landriau}, {Liu}, {Niemeyer},
  {MacQueen}, {Marshall}, {Mawatari}, {McLinden}, {Mukae}, {Nagaraj}, {Ono},
  {Ouchi}, {Papovich}, {Sakai}, {Saito}, {Schneider}, {Schulze},
  {Shanmugasundararaj}, {Shetrone}, {Sneden}, {Snigula}, {Steinmetz}, {Thomas},
  {Thomas}, {Tuttle}, {Urrutia}, {Wisotzki}, {Wold}, {Zeimann}, \&
  {Zhang}}]{Gebhardt2021}
{Gebhardt}, K., {Mentuch Cooper}, E., {Ciardullo}, R., {et~al.} 2021, \apj,
  923, 217

\bibitem[{{Guo} {et~al.}(2020){Guo}, {Maiolino}, {Jiang}, {Matsuoka}, {Nagao},
  {Dors}, {Ginolfi}, {Henden}, {Bennett}, {Sijacki}, \& {Puchwein}}]{Guo2020}
{Guo}, Y., {Maiolino}, R., {Jiang}, L., {et~al.} 2020, \apj, 898, 26

\bibitem[{{Haiman} {et~al.}(2009){Haiman}, {Kocsis}, \& {Menou}}]{Haiman2009}
{Haiman}, Z., {Kocsis}, B., \& {Menou}, K. 2009, \apj, 700, 1952

\bibitem[{{Hamann} {et~al.}(2017){Hamann}, {Zakamska}, {Ross}, {Paris},
  {Alexandroff}, {Villforth}, {Richards}, {Herbst}, {Brandt}, {Cook}, {Denney},
  {Greene}, {Schneider}, \& {Strauss}}]{Hamann2017}
{Hamann}, F., {Zakamska}, N.~L., {Ross}, N., {et~al.} 2017, \mnras, 464, 3431

\bibitem[{{Hill} {et~al.}(2021){Hill}, {Lee}, {MacQueen}, {Kelz}, {Drory},
  {Vattiat}, {Good}, {Ramsey}, {Kriel}, {Peterson}, {DePoy}, {Gebhardt},
  {Marshall}, {Tuttle}, {Bauer}, {Chonis}, {Fabricius}, {Froning},
  {H{\"a}user}, {Indahl}, {Jahn}, {Landriau}, {Leck}, {Montesano}, {Prochaska},
  {Snigula}, {Zeimann}, {Bryant}, {Damm}, {Fowler}, {Janowiecki}, {Martin},
  {Mrozinski}, {Odewahn}, {Rostopchin}, {Shetrone}, {Spencer}, {Mentuch
  Cooper}, {Armandroff}, {Bender}, {Dalton}, {Hopp}, {Komatsu}, {Nicklas},
  {Ramsey}, {Roth}, {Schneider}, {Sneden}, \& {Steinmetz}}]{Hill2021}
{Hill}, G.~J., {Lee}, H., {MacQueen}, P.~J., {et~al.} 2021, \aj, 162, 298

\bibitem[{{Kormendy} \& {Ho}(2013)}]{Kormendy2013}
{Kormendy}, J., \& {Ho}, L.~C. 2013, \araa, 51, 511

\bibitem[{{Lau} {et~al.}(2022){Lau}, {Hamann}, {Gillette}, {Perrotta}, {Rupke},
  {Wylezalek}, \& {Zakamska}}]{Lau2022}
{Lau}, M.~W., {Hamann}, F., {Gillette}, J., {et~al.} 2022, arXiv e-prints,
  arXiv:2203.06203

\bibitem[{{Liu} {et~al.}(2022){Liu}, {Gebhardt}, {Mentuch Cooper}, {Davis},
  {Schneider}, {Ciardullo}, {Farrow}, {Finkelstein}, {Gronwall}, {Guo}, {Hill},
  {House}, {Jeong}, {Jogee}, {Kollatschny}, {Krumpe}, {Landriau}, {Chavez
  Ortiz}, \& {Zhang}}]{Liu2022}
{Liu}, C., {Gebhardt}, K., {Mentuch Cooper}, E., {et~al.} 2022, arXiv e-prints,
  arXiv:2204.13658

\bibitem[{{Loeb}(2007)}]{Loeb2007}
{Loeb}, A. 2007, \prl, 99, 041103

\bibitem[{{Luo} {et~al.}(2017){Luo}, {Brandt}, {Xue}, {Lehmer}, {Alexander},
  {Bauer}, {Vito}, {Yang}, {Basu-Zych}, {Comastri}, {Gilli}, {Gu},
  {Hornschemeier}, {Koekemoer}, {Liu}, {Mainieri}, {Paolillo}, {Ranalli},
  {Rosati}, {Schneider}, {Shemmer}, {Smail}, {Sun}, {Tozzi}, {Vignali}, \&
  {Wang}}]{Luo2017}
{Luo}, B., {Brandt}, W.~N., {Xue}, Y.~Q., {et~al.} 2017, \apjs, 228, 2

\bibitem[{{McIntosh} {et~al.}(1999){McIntosh}, {Rieke}, {Rix}, {Foltz}, \&
  {Weymann}}]{McIntosh1999}
{McIntosh}, D.~H., {Rieke}, M.~J., {Rix}, H.~W., {Foltz}, C.~B., \& {Weymann},
  R.~J. 1999, \apj, 514, 40

\bibitem[{{McLure} {et~al.}(1999){McLure}, {Kukula}, {Dunlop}, {Baum}, {O'Dea},
  \& {Hughes}}]{McLure1999}
{McLure}, R.~J., {Kukula}, M.~J., {Dunlop}, J.~S., {et~al.} 1999, \mnras, 308,
  377

\bibitem[{{Ouchi} {et~al.}(2020){Ouchi}, {Ono}, \& {Shibuya}}]{Ouchi2020}
{Ouchi}, M., {Ono}, Y., \& {Shibuya}, T. 2020, \araa, 58, 617

\bibitem[{{P{\^a}ris} {et~al.}(2018){P{\^a}ris}, {Petitjean}, {Aubourg},
  {Myers}, {Streblyanska}, {Lyke}, {Anderson}, {Armengaud}, {Bautista},
  {Blanton}, {Blomqvist}, {Brinkmann}, {Brownstein}, {Brandt}, {Burtin},
  {Dawson}, {de la Torre}, {Georgakakis}, {Gil-Mar{\'\i}n}, {Green}, {Hall},
  {Kneib}, {LaMassa}, {Le Goff}, {MacLeod}, {Mariappan}, {McGreer}, {Merloni},
  {Noterdaeme}, {Palanque-Delabrouille}, {Percival}, {Ross}, {Rossi},
  {Schneider}, {Seo}, {Tojeiro}, {Weaver}, {Weijmans}, {Y{\`e}che}, {Zarrouk},
  \& {Zhao}}]{Paris2018}
{P{\^a}ris}, I., {Petitjean}, P., {Aubourg}, {\'E}., {et~al.} 2018, \aap, 613,
  A51

\bibitem[{{Rakshit} {et~al.}(2020){Rakshit}, {Stalin}, \&
  {Kotilainen}}]{Rakshit2020}
{Rakshit}, S., {Stalin}, C.~S., \& {Kotilainen}, J. 2020, \apjs, 249, 17

\bibitem[{{Ricarte} {et~al.}(2021){Ricarte}, {Tremmel}, {Natarajan}, {Zimmer},
  \& {Quinn}}]{Ricarte2021}
{Ricarte}, A., {Tremmel}, M., {Natarajan}, P., {Zimmer}, C., \& {Quinn}, T.
  2021, \mnras, 503, 6098

\bibitem[{{Richards} {et~al.}(2006){Richards}, {Strauss}, {Fan}, {Hall},
  {Jester}, {Schneider}, {Vanden Berk}, {Stoughton}, {Anderson}, {Brunner},
  {Gray}, {Gunn}, {Ivezi{\'c}}, {Kirkland}, {Knapp}, {Loveday}, {Meiksin},
  {Pope}, {Szalay}, {Thakar}, {Yanny}, {York}, {Barentine}, {Brewington},
  {Brinkmann}, {Fukugita}, {Harvanek}, {Kent}, {Kleinman}, {Krzesi{\'n}ski},
  {Long}, {Lupton}, {Nash}, {Neilsen}, {Nitta}, {Schlegel}, \&
  {Snedden}}]{Richards2006}
{Richards}, G.~T., {Strauss}, M.~A., {Fan}, X., {et~al.} 2006, \aj, 131, 2766

\bibitem[{{Ross} {et~al.}(2015){Ross}, {Hamann}, {Zakamska}, {Richards},
  {Villforth}, {Strauss}, {Greene}, {Alexandroff}, {Brandt}, {Liu}, {Myers},
  {P{\^a}ris}, \& {Schneider}}]{Ross2015}
{Ross}, N.~P., {Hamann}, F., {Zakamska}, N.~L., {et~al.} 2015, \mnras, 453,
  3932

\bibitem[{{Shimwell} {et~al.}(2022){Shimwell}, {Hardcastle}, {Tasse}, {Best},
  {R{\"o}ttgering}, {Williams}, {Botteon}, {Drabent}, {Mechev}, {Shulevski},
  {van Weeren}, {Bester}, {Br{\"u}ggen}, {Brunetti}, {Callingham}, {Chy{\.z}y},
  {Conway}, {Dijkema}, {Duncan}, {de Gasperin}, {Hale}, {Haverkorn}, {Hugo},
  {Jackson}, {Mevius}, {Miley}, {Morabito}, {Morganti}, {Offringa}, {Oonk},
  {Rafferty}, {Sabater}, {Smith}, {Schwarz}, {Smirnov}, {O'Sullivan},
  {Vedantham}, {White}, {Albert}, {Alegre}, {Asabere}, {Bacon}, {Bonafede},
  {Bonnassieux}, {Brienza}, {Bilicki}, {Bonato}, {Calistro Rivera}, {Cassano},
  {Cochrane}, {Croston}, {Cuciti}, {Dallacasa}, {Danezi}, {Dettmar}, {Di
  Gennaro}, {Edler}, {En{\ss}lin}, {Emig}, {Franzen}, {Garc{\'\i}a-Vergara},
  {Grange}, {G{\"u}rkan}, {Hajduk}, {Heald}, {Heesen}, {Hoang}, {Hoeft},
  {Horellou}, {Iacobelli}, {Jamrozy}, {Jeli{\'c}}, {Kondapally}, {Kukreti},
  {Kunert-Bajraszewska}, {Magliocchetti}, {Mahatma}, {Ma{\l}ek}, {Mandal},
  {Massaro}, {Meyer-Zhao}, {Mingo}, {Mostert}, {Nair}, {Nakoneczny},
  {Nikiel-Wroczy{\'n}ski}, {Orr{\'u}}, {Pajdosz-{\'S}mierciak}, {Pasini},
  {Prandoni}, {van Piggelen}, {Rajpurohit}, {Retana-Montenegro}, {Riseley},
  {Rowlinson}, {Saxena}, {Schrijvers}, {Sweijen}, {Siewert}, {Timmerman},
  {Vaccari}, {Vink}, {West}, {Wo{\l}owska}, {Zhang}, \& {Zheng}}]{Shimwell2022}
{Shimwell}, T.~W., {Hardcastle}, M.~J., {Tasse}, C., {et~al.} 2022, \aap, 659,
  A1

\bibitem[{{Stern} {et~al.}(2012){Stern}, {Assef}, {Benford}, {Blain}, {Cutri},
  {Dey}, {Eisenhardt}, {Griffith}, {Jarrett}, {Lake}, {Masci}, {Petty},
  {Stanford}, {Tsai}, {Wright}, {Yan}, {Harrison}, \& {Madsen}}]{Stern2012}
{Stern}, D., {Assef}, R.~J., {Benford}, D.~J., {et~al.} 2012, \apj, 753, 30

\bibitem[{{Vayner} {et~al.}(2021){Vayner}, {Zakamska}, {Riffel}, {Alexandroff},
  {Cosens}, {Hamann}, {Perrotta}, {Rupke}, {Bergmann}, {Veilleux}, {Walth},
  {Wright}, \& {Wylezalek}}]{Vayner2021}
{Vayner}, A., {Zakamska}, N.~L., {Riffel}, R.~A., {et~al.} 2021, \mnras, 504,
  4445

\bibitem[{{Xue} {et~al.}(2016){Xue}, {Luo}, {Brandt}, {Alexander}, {Bauer},
  {Lehmer}, \& {Yang}}]{Xue2016}
{Xue}, Y.~Q., {Luo}, B., {Brandt}, W.~N., {et~al.} 2016, \apjs, 224, 15

\end{thebibliography}

\end{document}